\documentclass[%
 reprint,
superscriptaddress,
 amsmath,amssymb,
 aps,
]{revtex4-2}

\usepackage{amsmath}    
\usepackage{graphicx}   
\usepackage{verbatim}   
\usepackage{color}      
\usepackage{subfigure}  
\usepackage{hyperref}   
\usepackage{epstopdf}
\usepackage{color}

\usepackage{hyperref}
\providecommand{\keywords}[1]{\textbf{Keywords:} #1}

\begin{document}

\title{Long-Range Anomalous Decay of the Correlation in Jammed Packings}

\author{Paolo Rissone}
\email{paolo.rissone@ub.edu}
\affiliation{Small Biosystems Lab, Department of Condensed Matter Physics, Carrer de Marti i Franques, 1, 11, 08028, Barcelona, Spain}

\author{Eric I. Corwin}
\email{ecorwin@uoregon.edu}
\affiliation{Department of Physics and Materials Science Institute, University of Oregon, Eugene, Oregon 97403, USA}

\author{Giorgio Parisi}
\email{giorgio.parisi@roma1.infn.it}
\affiliation{Dipartimento di Fisica, Sapienza Università di Roma, P.le Aldo Moro 5, 00185 Rome, Italy}
\affiliation{Istituto Nazionale di Fisica Nucleare, Sezione di Roma I, P.le A. Moro 5, 00185 Rome, Italy}
\affiliation{Institute of Nanotechnology (NANOTEC) - CNR, Rome unit, P.le A. Moro 5, 00185 Rome, Italy}

\date{\today}

\begin{abstract}
We numerically study the structure of the interactions occurring in three-dimensional systems of hard spheres at jamming, focusing on the large-scale behavior. Given the fundamental role they play in the configuration of jammed packings, we analyze the propagation through the system of the weak forces and of the variation of the coordination number with respect to the isostaticity condition $\Delta Z$. We show that these correlations can be successfully probed by introducing a correlation function weighted on the density-density fluctuations. The results of this analysis can be further improved by introducing a representation of the system based on the contact points between particles. In particular, we find evidence that the weak forces and the $\Delta Z$ fluctuations support the hypothesis of randomly jammed packings of spherical particles being hyperuniform by exhibiting an anomalous long-range decay. Moreover, we find that the large-scale structure of the density-density correlation exhibits a complex behavior due to the superimposition of two exponentially damped oscillating signals propagating with linearly depending frequencies.
\end{abstract}

\maketitle
\cleardoublepage

\narrowtext

\textit{Introduction.} Amorphous packings of nearly incompressible particles, such as marbles and pebbles, have been the object of an intense investigation during the past decades as they represent a suitable benchmark for studying a broad range of dense-packing and optimization problems \cite{boettcher2002jamming,krzakala2007landscape}. This rising interest led to the development of many experiments~\cite{majmudar2007jamming,zhang2010jamming,behringer2014statistical} and simulations~\cite{bitzek2006structural,zhang2015structure,corwin2010model} that made possible an extensive study of the features of these systems. Moreover, this field appeared to be the perfect environment to apply the theories of frustrated interactions~\cite{charbonneau2013geometrical}. In particular, the application of the replica theory~\cite{parisi2009replica} led to the elaboration of an exact analytical solution valid in the limit of high-dimensional packings~\cite{kurchan2012exact, kurchan2013exact, charbonneau2014exact}.

We focus on athermal packings of frictionless hard spheres (HSs) compressed until particles come into mechanical contact with their nearest neighbors. The trapped spheres form a rigid network and cannot explore the surrounding environment (ergodicity breaking). Under these conditions, the system enters a phase of matter known as ``jamming"~\cite{biroli2007jamming, charbonneau2016glass}. It has been hypothesized that saturated jammed systems (no space to add another particle) are hyperuniform~\cite{torquato2003local}, implying that their radial distribution function (RDF) tends to zero from negative values and as a power law~\cite{donev2005pair} -- $g(r)-1 \propto r^{-4}$. Even though the tendency of the jammed packings to hyperuniformity has been observed (with deviations from the postulated behavior)~\cite{ikeda2017large}, such power-law scaling of the pair correlation function has never been directly measured.

In this framework, we find evidence of hyperuniformity in the long-range correlation of the forces exchanged between adjacent particles and the deviation of the number of contacts per particle from the average value, $\Delta Z = Z - \langle Z\rangle$. On the one hand, jammed packings exhibit a unique force network~\cite{charbonneau2015jamming}, whose long-range fluctuations demand study. On the other hand, it has been shown that $\Delta Z$ exhibits interesting features at jamming~\cite{wyart2005effects,karayiannis2009contact,moukarzel2012elastic} and that the fluctuations of the coordination number $\sigma_Z^2$ for a fixed $\Delta Z$ are similar to those of density hyperuniformity~\cite{hexner2018diverging}.
The research for a static observable exhibiting a nontrivial behavior close to jamming is motivated by the existence of a corresponding long-ranged dynamical response. It has been shown~\cite{degiuli2015theory, muller2015marginal, wyart2005effects} that a local perturbation to the position of a pair of adjacent particles, i.e., breaking the contact between particles, produces a response propagating through the system up to a maximum length, the 
``response length'', $\xi_R$ that diverges at jamming~\cite{liu2010jamming}.
The main hindrance to this analysis is represented by the strong statistical noise exhibited by the correlation functions in the long range and superimposing to the (weak) signal of interest. To overcome this problem, we define a suitable pair correlation function to point out the long-range behavior of the observables by filtering out the interfering signals. Moreover, we introduce a representation of the interparticle network based on the contact points between particles instead of their centers of mass. We show that the shift to a system of fictive particles improves the resolution of the correlation function and is fundamental in identifying the long-range features of the jammed packings.

%

\textit{System Setup.} Given a system of $N$ randomly distributed monodisperse HSs of diameter $\sigma$, let us introduce the interaction via the dimensionless interaction potential
\begin{equation}
\label{eq:HSpotential}
U=\sum_{\langle i,j \rangle} \left( 1 - \frac{|\mathbf{r}_i - \mathbf{r}_j|}{\sigma_{i j}} \right) \, ,
\end{equation}
where $\mathbf{r}_{i,j}$ is the position of particles $i,j$ and $\sigma_{ij}$ is the distance between the centers of particles $i$ and $j$ when they are in contact. Note that, for monodisperse HSs, $\sigma_{ij}=\sigma$ and that for particles in kissing contact $r = |\mathbf{r}_i - \mathbf{r}_j| = \sigma$ so that particles' interaction potential is zero.

The system is controlled via the ``packing fraction'' $\phi$, defined as the fraction of the system volume occupied by the spheres. The jammed phase is reached when the packing fraction hits the critical value $\phi_J \approx 0.64$ in three-dimensional systems~\cite{donev2005neighbor,kim2003glass,silbert2006structural,somfai2007critical}. By starting in the overjammed region $(\phi \approx 2\phi_J)$, $\phi$ is gradually decreased by gently shrinking the particles' diameter until the system reaches jamming. The jamming point is approached by iteratively minimizing the potential energy [Eq. \eqref{eq:HSpotential}] using the 
``FIRE'' algorithm~\cite{bitzek2006structural} according to the protocol described in~\cite{charbonneau2015jamming} (Appendix B). The simulation ends when the targeted precision is reached, i.e. when it is impossible to distinguish between a kissing contact and a small overlap. 

At the end of this protocol, the particles form a network of enduring contacts that is stable only if the isostaticity condition holds~\cite{wyart2005effects,van2009jamming,torquato2010jammed,coulais2012dense}, i.e., the average number of contacts per particle satisfies (in a first approximation) $\langle Z \rangle \equiv Z_{iso} \approx 2d$, where $d$ is the system dimension. 
As shown in \cite{zaccone2011approximate}, this property depends on the system mechanical stability being ensured by the exact balancing between the affine and the nonaffine (negative) components of the interparticle interactions over the whole network.
Within this picture, it is important to point out the existence of ``bucklers'', i.e., particles that still form part of the rigid network but that are minimally constrained so that $Z = d+1$~\cite{charbonneau2015jamming}. A notable exception to the isostaticity condition is represented by ``rattlers'', particles that are not part of the contact network and freely move inside cages bounded by particles permanently in contact. The identification and the exclusion of rattlers are fundamental for obtaining reliable results. By adopting this method, we generated 96 critically jammed packings of $N=16\,384$ particles in $d=3$.
%

\textit{Generalized RDF.} Given a generic observable $\mathcal{O}$, we defined the ``generalized pair correlation function'' as 
\begin{equation}
\label{eq:corr_generic}
C^s_{\mathcal{O}}(r) = \frac{g^s_{\mathcal{O}}(r)}{g(r)} \, ,
\end{equation}
where $r$ is the distance between particle pairs, $g(r)$ is the usual RDF~\cite{frenkel2001understanding} and 
\begin{equation}
\label{eq:rdf_generic}
g^s_{\mathcal{O}}(r)= \frac{1}{C} \sum_{i,j} \delta(|\mathbf{r}_i - \mathbf{r}_j| - r) \mathcal{O}^s_i \mathcal{O}^s_j \, ,
\end{equation}
$\mathbf{r}_i$ and $\mathbf{r}_j$ being the positions of particles $i$ and $j$, $C$ the normalization factor and $s \in \Re$ a control parameter. Notice that by choosing $s=0$ in Eq. \eqref{eq:rdf_generic}, one gets the RDF (additional details can be found in the Supplemental Material\footnote{See Supplemental Material for the extensive description of the generalized RDF}). We studied the two cases $\mathcal{O} = f$ and $\mathcal{O} = \Delta Z \equiv Z - Z_{\rm iso}$, $f$ being the force exchanged between particles and $\Delta Z$ the deviation of the number of contacts per particle from isostaticity.

\begin{figure}[ht]
\centering
\includegraphics[width=0.43\textwidth]{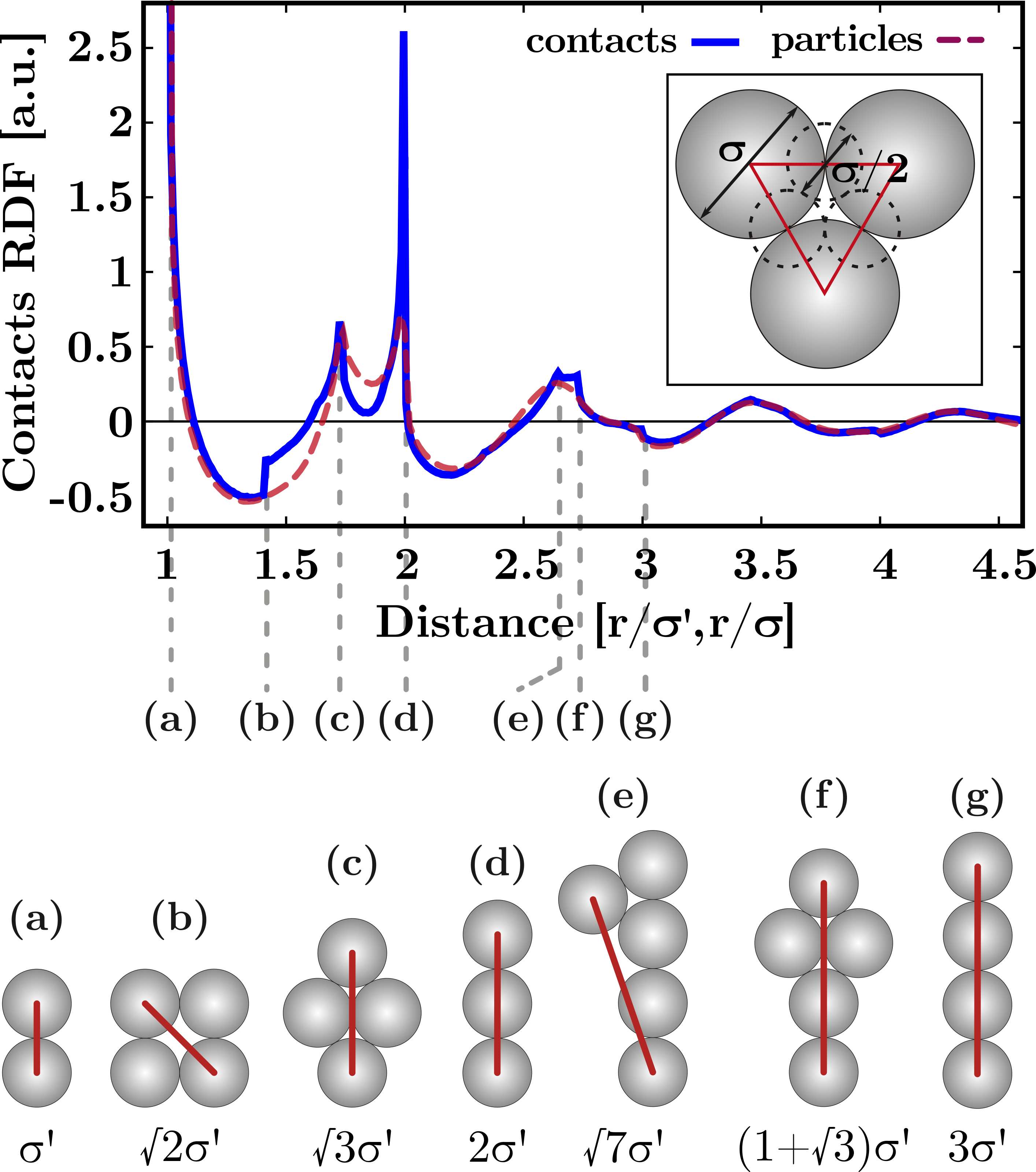}
\caption{\label{fig:cartoon_rdf} Contacts and original RDFs (in blue and red, respectively) in $d=3$ as a function of the distance (expressed in the respective diameter units $r/\sigma'$ and $r/\sigma$). Each peak (labeled with a different letter) occurs in correspondence to a different configuration of the fictive particles, as shown in the correspondent representations (which, however, do not exhaust all the possibilities), along with the distance at which each discontinuity appears (red lines). \textbf{(Inset)} Schematic representation of the contacts-centered spheres model in $d=3$. 
Each fictive particle (dashed black lines) results from the contact of two HSs and has a diameter $\sigma'=\sigma/2$, being $\sigma$ the diameter of the original HS.}
\end{figure}

To study the long-range correlations with higher accuracy, we also introduced the jammed packings' representation with respect to the contact points between particles instead of the centers of mass. To do this, let us consider the densest packing in which HSs can arrange (see the inset in Fig. \ref{fig:cartoon_rdf}). Each contact point between the HSs can be seen as the center of a fictive particle with radius $\sigma' = \sigma/2$. Thus, given a system of $N$ particles and $Z_{\rm iso}=2d$ average contacts per particle, the new contacts-based system will be formed by $N' \leq 2dN = 6N$ particles.
%

\textit{Contacts RDF.} Switching from the centers-of-mass-based to the contact-points description of the network, allowed us to study the radial distribution function of the contact-centered model (called ``contacts RDF'' in what follows) with a much higher resolution. 
Figure \ref{fig:cartoon_rdf} shows the short-range behavior of the contacts RDF (blue line) compared to the real spheres RDF (red dashed line).
The contacts RDF points out new features of the pair correlation function at jamming, evidencing discontinuities that were much smoother [peaks (e),(f),(g)] or completely absent [peak (b)] in the original system description. The higher accuracy of the contacts RDF is further reflected by the enhanced sharpness of peaks (c) and (d). The origin of each one of these discontinuities can be easily addressed. The first peak ($r=\sigma'$) is due to a nearest-neighbor contact, while the others are determined by different possible configurations of HSs forming a chain of contacts. Notably, peaks (c) and (d) in Fig.\ref{fig:cartoon_rdf} correspond to ``real'' $\delta$ functions, i.e., to mechanically rigid configurations of perfectly centrosymmetric particles with respect to the transmission of forces. The local centrosymmetry ensures a zero nonaffine component of the response and therefore the full mechanical stability \cite{milkus2016local}. All the other peaks are the result of a wide range of possible arrangements.
\begin{figure}[ht]
\centering
\includegraphics[width=0.42\textwidth]{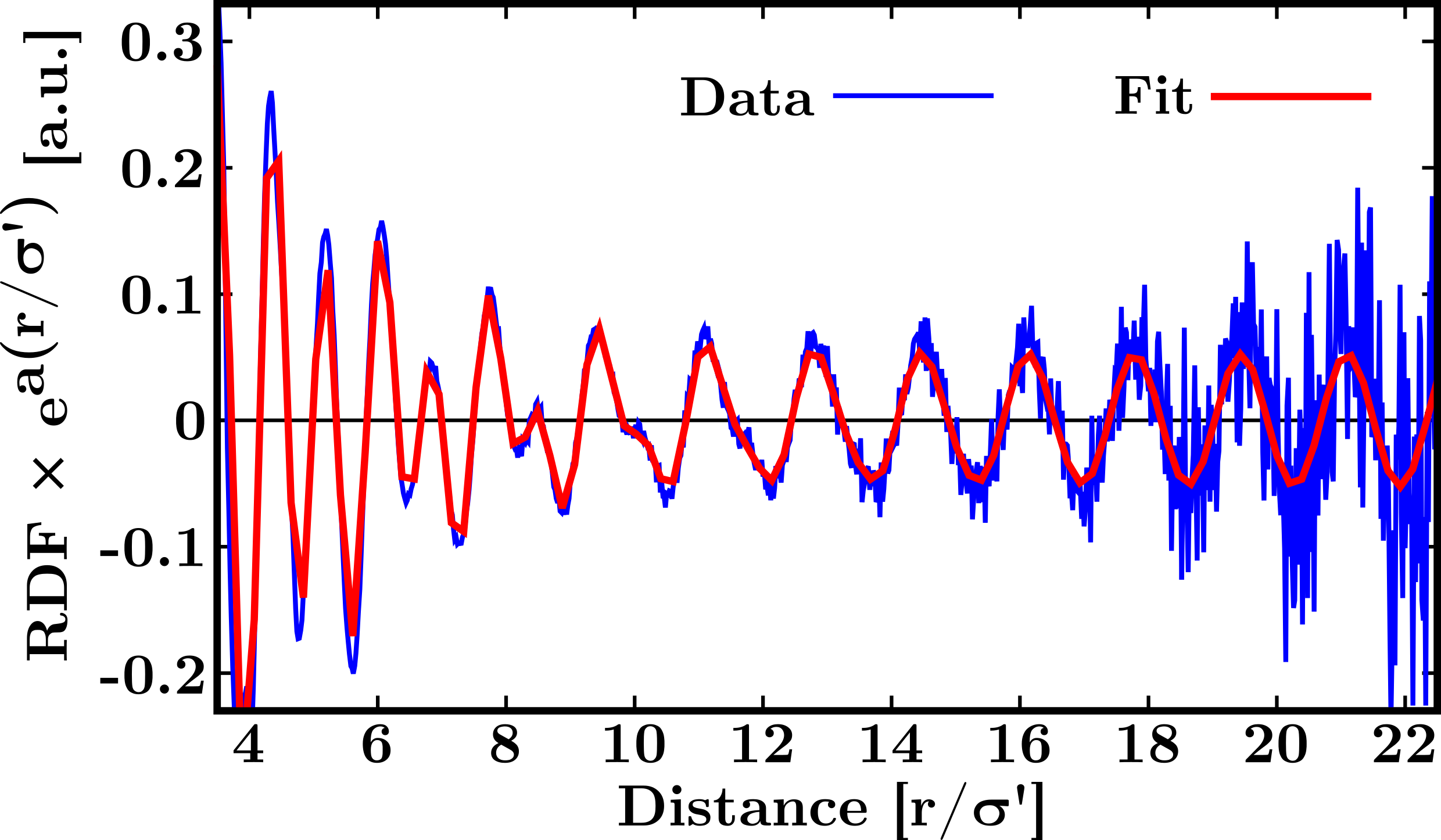}
\caption{ \label{fig:3d_contacts_fit8parameters} Contacts density-density correlation $g(r/\sigma')$ multiplied by the factor $e^{a\, r/\sigma'}$ with $a\approx 0.3$. Two sets of distinct oscillations given by Eq. \eqref{eq:fit_f} appear to be superimposed. The first one, $f_m(r)$, propagating in the middle range $[4:9] \, r/\sigma '$ and the second, $f_l(r)$, in the long-range region $[9:22] \, r/\sigma '$. By fitting the resulting function (red line) we found that $p_l\approx 7.5$ and $p_m\approx3.8$. The distance is expressed in diameter units.}
\end{figure}
Interestingly, as shown in Fig.\ref{fig:3d_contacts_fit8parameters}, the mid-range and long-range behavior of the correlation function appears to be exponentially damped as $e^{-ar}$, with $a\approx0.3$. The enhancement of such behavior by considering $g(r) \times e^{ar}$ pointed out a superposition of two oscillatory functions of type
\begin{equation}
\label{eq:fit_f}
f_i(r)= c_i \, e^{a_i r} \cos(p_i r + \psi_i),
\end{equation}
where $c_i, a_i, p_i, \psi_i$ are the function parameters and $i=m,l$ denotes the mid- and long-range regions, respectively. By performing the resulting eight parameters fit\footnote{See Supplemental Material for a complete description of the 8 parameters fit} $g(r) = f_m(r) + f_l(r)$, we found $p_m \approx 2p_l = 7.514 \pm 0.003$ and $a_m \approx 2a_l =  0.70 \pm 0.01$. This result proves that because the contribution of the long-range oscillations is small compared to the mid-range ones, the $f_l(r)$ can be considered as $\mathcal{O}(2)$ correction to the $\mathcal{O}(1)$ mid-range leading term $f_m(r)$. 
%

\textit{Weak Forces Correlation.} At the jamming point, each particle of the system gets trapped in a fixed position by mechanical contact with its nearest neighbors. The stability of the resulting configuration is ensured by the balance of all the forces exchanged in these contact points, determining the formation of a complex forces network spreading through the whole system \cite{wang2018microscopic,hartley2003logarithmic,behringer2014statistical}. Within this picture, it is possible to distinguish between 
``strong'' forces, which form a backbone crossing the whole system, and ``weak'' forces spreading only in small subregions of the system confined by branches of the main network (see Supplemental Material\footnote{See the Supplemental Material for a detailed explanation on origin of the weak forces}). 
Let us consider Eq. \eqref{eq:rdf_generic} with $\mathcal{O}=f$. By choosing $s<0$ the resulting $g_f^s(r)$ will be ``weighted'' on the weak forces so that the smaller the force, the bigger its contribution to the correlation. Note that not \textit{any} value of $s$ can be chosen. In fact, as shown in \cite{charbonneau2015jamming}, the force distribution can be described as function $P(f) \propto f^{\theta}$, with $\theta \approx 0.4$, Thus, the average force (raised to the power $s$) can be estimated as
\begin{equation}
\label{eq:force_distribution}
\langle f^s \rangle \sim \int df f^s P(f) = \int df f^s f^{\theta} \propto \frac{1}{1+\theta+s} \, .
\end{equation}
which diverges unless $s>s_{\min} \equiv -1-\theta \approx -1.4$.
\begin{figure}[ht]
    \centering
    \subfigure{\label{fig:forces_s=-1}
        \includegraphics[width=0.4\textwidth]{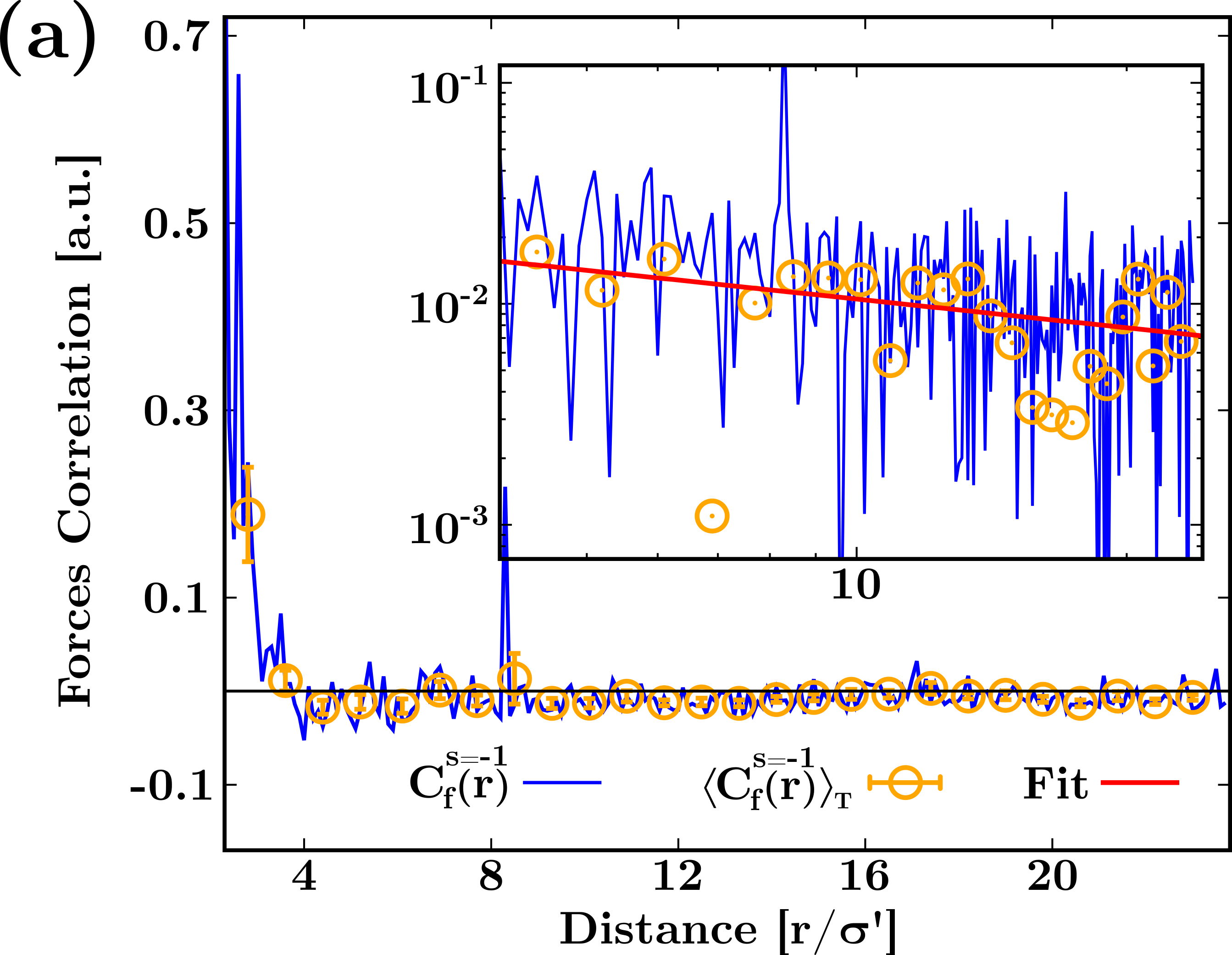}}
    \subfigure{\label{fig:forces_s=-0.5}
        \includegraphics[width=0.4\textwidth]{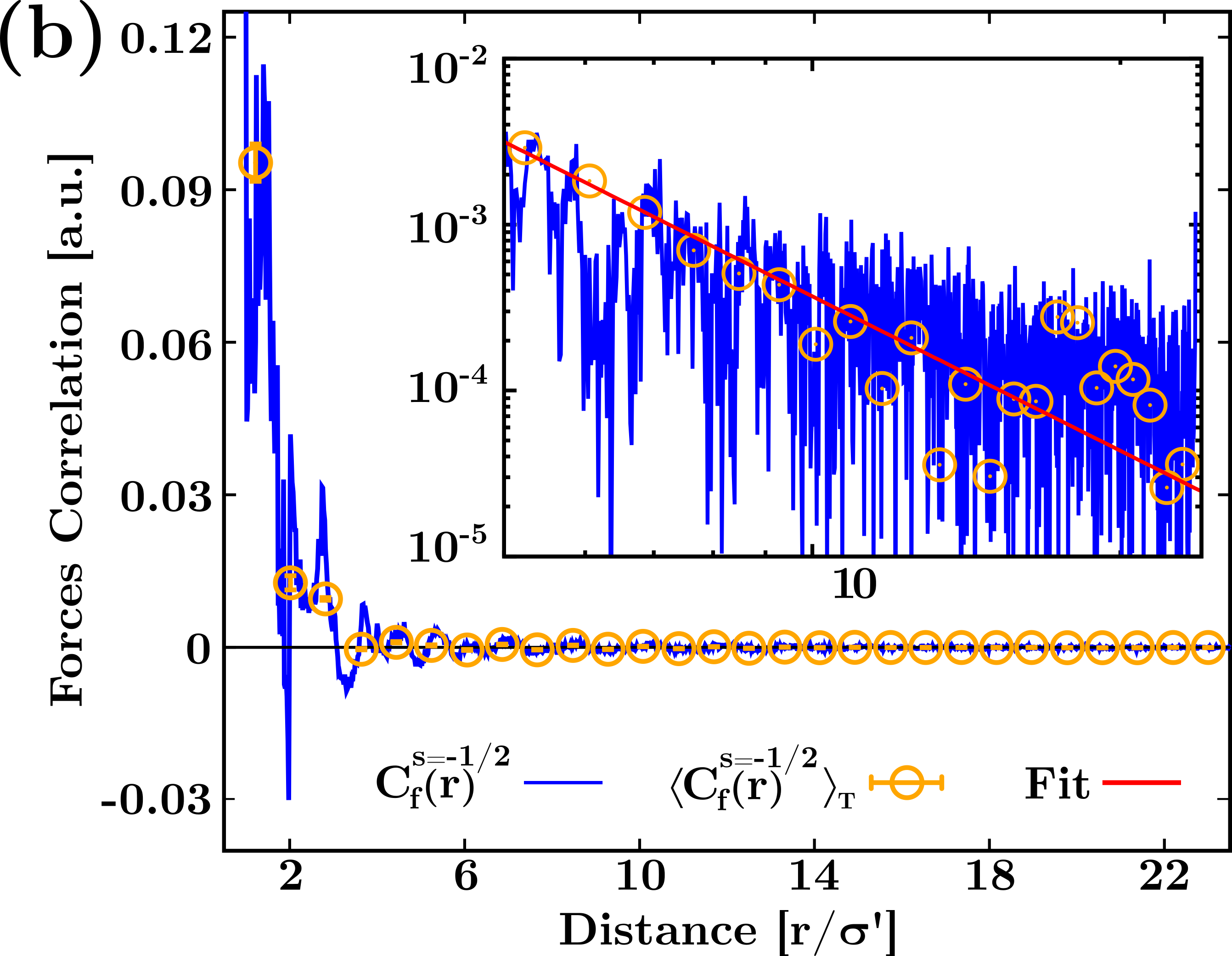}}
    \caption{\label{fig:weak_forces} \textbf{(a)} Weak force correlation function $C^{s=-1}_f(r)$ in $d=3$ and enlargement of the long-range region in log-log scale (inset). The density-density correlations propagating at long range have been filtered by introducing the average over the oscillation period $T$, $\langle C^{s}_f(r) \rangle _T$ (orange circles). The log-log scale in the inset points out a long-range power-law behavior (red line) decaying with an exponent $\gamma_f^{s=-1} = 0.7 \pm 0.3$ in the range [7:23].
    \textbf{(b)} Weak forces correlation function $C^{s=-1/2}_f(r)$ in $d=3$. The fluctuations analysis points out a power-law decay with an exponent $\gamma_f^{s=-1/2} = 4.1 \pm 0.3$ in the range [7:23]. Notice that both insets show the modulus of the correlation function. The distances are expressed in diameter units.}
\end{figure}
Figure \ref{fig:weak_forces} shows the weak forces correlation $C^s_f(r)$ for $s=-1$ and $s=-1/2$. In both cases, the local density-density oscillations appear to be not completely damped, propagating up to large length scales and masking any power-law decay. Therefore, we introduced the correlation function averaged over the period $T$ of the short-range oscillations $\langle C^{s}_f(r) \rangle _T$ (orange circles), filtering out most of the additional periodic component. The period $T$ is roughly equal to the particle diameter but is tuned for each dataset. By fitting in the mid- and long-range region $[7:23] \, r/\sigma'$ and the resulting correlation functions to the power law $f(r) = A r^{-\gamma_f^s} + C$ for $s=(-1,-1/2)$ (red line in the insets) with $C \approx 0$, we respectively found $\gamma_f^{s=-1} = 0.7 \pm 0.3$ and $\gamma_f^{s=-1/2} = 4.1 \pm 0.3$. These results prove that the fluctuations of the weak force correlation function at large length scales decay with the power law expected by the hyperuniformity theory, with the only constraint being a fine tuning of the selected (weak) forces. In fact, for $s=-1$ the fluctuations rapidly go to zero in the short range, whereas for $s=-1/2$ they exhibit a long-ranged anomalous decay with the measured nontrivial exponent.
%

\textit{Contacts Correlation.} It has been proved that, at jamming, the coordination number per particle -- $Z$ -- plays an essential role in determining the system features~\cite{hexner2018diverging}. The study of the fluctuations $\sigma_{\Delta Z}^2$ led to the definition of a structural relaxation length $\xi_{\Delta Z} \propto \Delta Z^{-\nu}$, where $\Delta Z = Z - 2d$, $d$ is the system dimension and $\nu$ is a nontrivial exponent.
Here we point out that the $\Delta Z$ correlation exhibits the expected anomalous behavior. 

\begin{figure}[ht]
\centering
\includegraphics[width=0.42\textwidth]{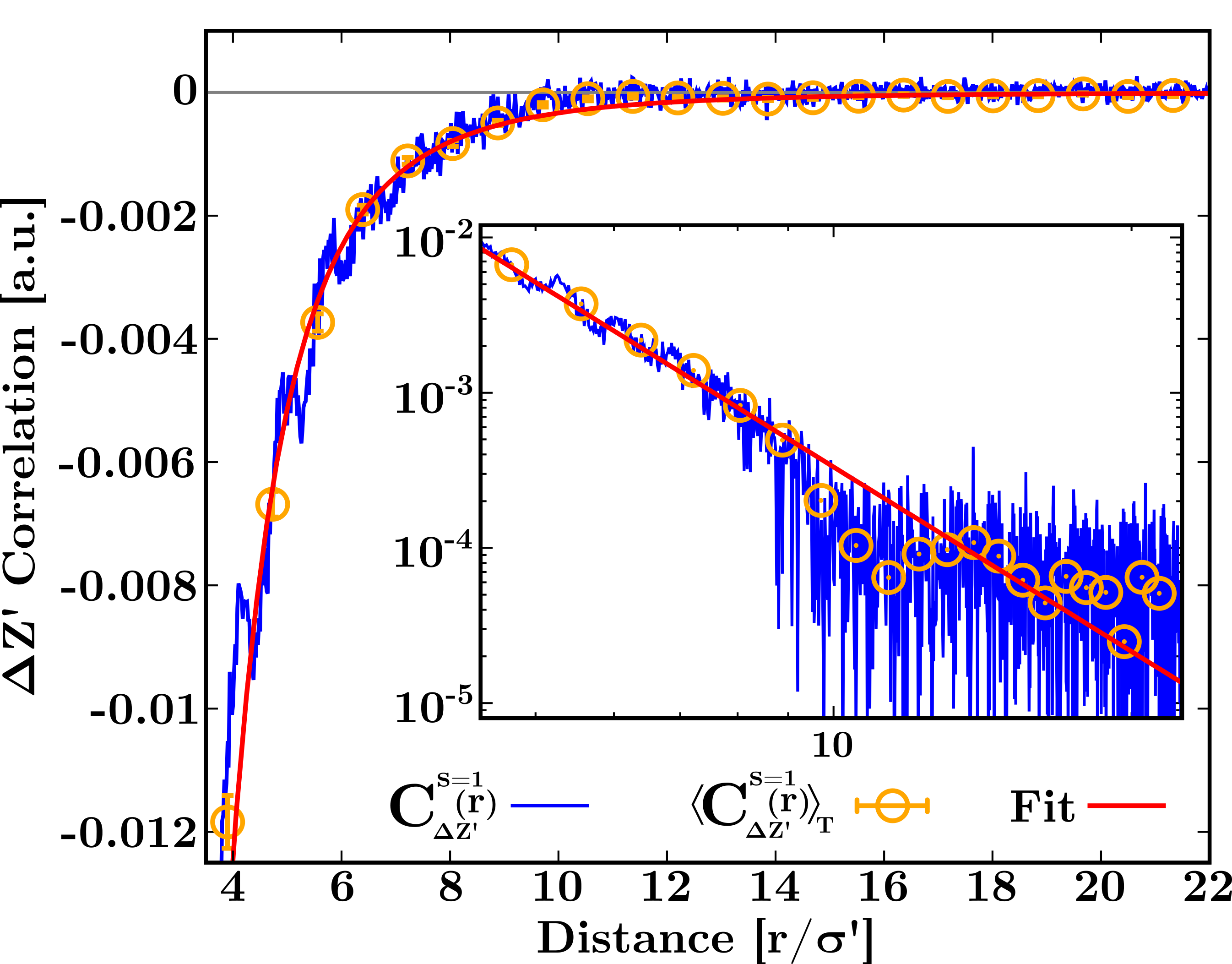}
\caption{ \label{fig:power_law} Contact correlation function $C^{s=1}_{\Delta Z'} (r)$ in $d=3$. The density-density oscillations at short range (blue line) have been filtered by introducing the correlation averaged over the oscillation period $T$, $\langle C_{\Delta Z'}^{s=1}(r) \rangle$ (orange circles).
(\textbf{Inset}) Enlargement of the mid- and long-range regions of the modulus of the $C^{s=1}_{\Delta Z'}(r)$ (blue line) in log-log scale. The correlation function exhibits a power-law decay (red line) propagating with a nontrivial exponent $\gamma_{\Delta Z'} = 3.9 \pm 0.2$. The distance is expressed in diameter units.}
\end{figure}

As previously discussed, changing to a contacts-based system representation also implies changing from a system of $N$ particles to one of $N' = 2dN$ fictive particles. In this framework, we defined the number of contacts per (fictive) particle as 
\begin{equation}
    Z'_l = \frac{Z_i + Z_j}{2} \, ,
\end{equation}
where $l=(1,\dots,N')$ and $i,j=(1,\dots,N)$ with $i\neq j$. Thus, we chose $\mathcal{O}=\Delta Z' = Z'-2d$ and $s=1$ Eqs. \eqref{eq:rdf_generic} and \eqref{eq:corr_generic}.
The resulting correlation function $C^{s=1}_{\Delta Z'} (r)$ (shown in Fig.\ref{fig:power_law}) clearly exhibits a nontrivial behavior in the middle and long-range superimposed onto some oscillations as a result of the incomplete damping of the density-density correlations. As described above, this power-low trend can be seen by fitting the period $T$ of the density-density oscillations and eventually recomputing Eq. \eqref{eq:rdf_generic} by considering spherical shells of thickness $T$ (orange circles in Fig.\ref{fig:power_law}). Analogously to the previous case, we fit the resulting function according to the power law $f(r) = Ar^{-\gamma_{\Delta Z'}} + C$ in the range [4:23], finding a nontrivial exponent $\gamma_{\Delta Z'} = 3.9 \pm 0.2$.
%

\textit{Conclusion.} We studied the mid- and large-scale spatial correlations in HS packings at jamming by defining the correlation functions [Eqs. \eqref{eq:rdf_generic} and \eqref{eq:corr_generic}] for the weak forces $f$ and the variation of the coordination number per particle $\Delta Z'$. We found that the correlation functions exhibit a long-range anomalous behavior, supporting the hyperuniformity hypothesis and marking the transition to the jammed phase. In fact, both the weak forces and the $\Delta Z'$ correlation exhibit a power-law decay with nontrivial exponents $\gamma^{s=-1}_f \approx \gamma_{\Delta Z'} \approx 4$. Moreover, we introduced a representation of the system based on the particles' contact points instead of their centers of mass. This change pointed out new features of the density-density correlation function, which appears to be a sum of two different oscillating signals propagating in the middle and long-range. A theoretical explanation for this two-terms form could be recovered by writing integral equations of the hypernetted chain kind \cite{hansen1990theory} for the correlation function. However this problem has not been explored yet. It would be of great interest to obtain such proof. Changing to the contacts-based representation of the system also increased the sensitivity of the correlation function at long range. 

These results show that static observables other than the density can be used to characterize the main features of jammed systems, marking (expected) theoretical behaviors that remained hidden by studying the density fluctuations of the pair correlation function. A further step in this analysis would involve the study of bigger systems, i.e. systems with a larger number of particles $N$, implying easier isolation of the power-law decay with respect to the local oscillations. Moreover, it would be of great interest to characterize these correlation functions at a packing fraction $\phi < \phi_J$ close to jamming and varying it up to the jamming point.
%

This work was supported by a Simons Foundation grant ($\#454939$, EC; $\#454949$, GP).


\bibliographystyle{ieeetr}
\bibliography{References.bib}

\begin{thebibliography}{10}

\bibitem{boettcher2002jamming}
S.~Boettcher and M.~Grigni, ``Jamming model for the extremal optimization
  heuristic,'' {\em J. Phys. A}, vol.~35, no.~5, p.~1109, 2002.

\bibitem{krzakala2007landscape}
F.~Krzakala and J.~Kurchan, ``Landscape analysis of constraint satisfaction
  problems,'' {\em Phys. Rev. E}, vol.~76, no.~2, p.~021122, 2007.

\bibitem{majmudar2007jamming}
T.~Majmudar, M.~Sperl, S.~Luding, and R.~P. Behringer, ``Jamming transition in
  granular systems,'' {\em Phys. Rev. Lett.}, vol.~98, no.~5, p.~058001, 2007.

\bibitem{zhang2010jamming}
J.~Zhang, T.~Majmudar, M.~Sperl, and R.~Behringer, ``Jamming for a 2d granular
  material,'' {\em Soft Matter}, vol.~6, no.~13, pp.~2982--2991, 2010.

\bibitem{behringer2014statistical}
R.~Behringer, D.~Bi, B.~Chakraborty, A.~Clark, J.~Dijksman, J.~Ren, and
  J.~Zhang, ``Statistical properties of granular materials near jamming,'' {\em
  J. Stat. Mech.: Theory Exp.}, vol.~2014, no.~6, p.~P06004, 2014.

\bibitem{bitzek2006structural}
E.~Bitzek, P.~Koskinen, F.~G{\"a}hler, M.~Moseler, and P.~Gumbsch, ``Structural
  relaxation made simple,'' {\em Phys. Rev. Lett.}, p.~170201, 2006.

\bibitem{zhang2015structure}
C.~Zhang, C.~B. O'Donovan, E.~I. Corwin, F.~Cardinaux, T.~G. Mason, M.~E.
  M{\"o}bius, and F.~Scheffold, ``Structure of marginally jammed polydisperse
  packings of frictionless spheres,'' {\em Phys. Rev. E}, vol.~91, no.~3,
  p.~032302, 2015.

\bibitem{corwin2010model}
E.~I. Corwin, M.~Clusel, A.~O. Siemens, and J.~Bruji{\'c}, ``Model for random
  packing of polydisperse frictionless spheres,'' {\em Soft Matter}, vol.~6,
  no.~13, pp.~2949--2959, 2010.

\bibitem{charbonneau2013geometrical}
B.~Charbonneau, P.~Charbonneau, and G.~Tarjus, ``Geometrical frustration and
  static correlations in hard-sphere glass formers,'' {\em J. Chem. Phys.},
  vol.~138, no.~12, p.~12A515, 2013.

\bibitem{parisi2009replica}
G.~Parisi and F.~Zamponi, ``A replica approach to glassy hard spheres,'' {\em
  J. Stat. Mech.: Theory Exp.}, p.~P03026, 2009.

\bibitem{kurchan2012exact}
J.~Kurchan, G.~Parisi, and F.~Zamponi, ``Exact theory of dense amorphous hard
  spheres in high dimension i. the free energy,'' {\em J. Stat. Mech.: Theory
  Exp.}, p.~P10012, 2012.

\bibitem{kurchan2013exact}
J.~Kurchan, G.~Parisi, P.~Urbani, and F.~Zamponi, ``Exact theory of dense
  amorphous hard spheres in high dimension. ii. the high density regime and the
  gardner transition,'' {\em J. Phys. Chem. B}, p.~12979, 2013.

\bibitem{charbonneau2014exact}
P.~Charbonneau, J.~Kurchan, G.~Parisi, P.~Urbani, and F.~Zamponi, ``Exact
  theory of dense amorphous hard spheres in high dimension. iii. the full
  replica symmetry breaking solution,'' {\em J. Stat. Mech.: Theory Exp.},
  p.~P10009, 2014.

\bibitem{biroli2007jamming}
G.~Biroli, ``Jamming: a new kind of phase transition?,'' {\em Nat. Phys.},
  p.~222, 2007.

\bibitem{charbonneau2016glass}
P.~Charbonneau, J.~Kurchan, G.~Parisi, P.~Urbani, and F.~Zamponi, ``Glass and
  jamming transitions: from exact results to finite-dimensional descriptions,''
  {\em Annu. Rev. Condens. Matter. Phys.}, p.~265, 2016.

\bibitem{torquato2003local}
S.~Torquato and F.~H. Stillinger, ``Local density fluctuations,
  hyperuniformity, and order metrics,'' {\em Phys. Rev. E}, vol.~68, no.~4,
  p.~041113, 2003.

\bibitem{donev2005pair}
A.~Donev, S.~Torquato, and F.~Stillinger, ``Pair correlation function
  characteristics of nearly jammed disordered and ordered hard-sphere
  packings,'' {\em Phys. Rev. E}, p.~011105, 2005.

\bibitem{ikeda2017large}
A.~Ikeda, L.~Berthier, and G.~Parisi, ``Large-scale structure of randomly
  jammed spheres,'' {\em Phys. Rev. E}, p.~052125, 2017.

\bibitem{charbonneau2015jamming}
P.~Charbonneau, E.~Corwin, G.~Parisi, and F.~Zamponi, ``Jamming criticality
  revealed by removing localized buckling excitations,'' {\em Phys. Rev.
  Lett.}, p.~125504, 2015.

\bibitem{wyart2005effects}
M.~Wyart, L.~Silbert, S.~Nagel, and T.~Witten, ``Effects of compression on the
  vibrational modes of marginally jammed solids,'' {\em Phys. Rev. E},
  p.~051306, 2005.

\bibitem{karayiannis2009contact}
N.~C. Karayiannis, K.~Foteinopoulou, and M.~Laso, ``Contact network in nearly
  jammed disordered packings of hard-sphere chains,'' {\em Phys. Rev. E},
  vol.~80, no.~1, p.~011307, 2009.

\bibitem{moukarzel2012elastic}
C.~F. Moukarzel, ``Elastic collapse in disordered isostatic networks,'' {\em
  EPL}, vol.~97, no.~3, p.~36008, 2012.

\bibitem{hexner2018diverging}
D.~Hexner, A.~J. Liu, and S.~Nagel, ``A diverging length scale in the structure
  of jammed systems,'' {\em Bull. Am. Phys. Soc.}, 2018.

\bibitem{degiuli2015theory}
E.~Degiuli, E.~Lerner, and M.~Wyart, ``Theory of the jamming transition at
  finite temperature,'' {\em J. Chem. Phys.}, p.~164503, 2015.

\bibitem{muller2015marginal}
M.~M{\"u}ller and M.~Wyart, ``Marginal stability in structural, spin, and
  electron glasses,'' {\em Annu. Rev. Condens. Matter. Phys.}, p.~177, 2015.

\bibitem{liu2010jamming}
A.~J. Liu and S.~R. Nagel, ``The jamming transition and the marginally jammed
  solid,'' {\em Annu. Rev. Condens. Matter Phys.}, vol.~1, no.~1, pp.~347--369,
  2010.

\bibitem{donev2005neighbor}
A.~Donev, S.~Torquato, and F.~H. Stillinger, ``Neighbor list collision-driven
  molecular dynamics simulation for nonspherical hard particles. i. algorithmic
  details,'' {\em J. Comput. Phys.}, vol.~202, no.~2, pp.~737--764, 2005.

\bibitem{kim2003glass}
K.~Kim and T.~Munakata, ``Glass transition of hard sphere systems: Molecular
  dynamics and density functional theory,'' {\em Phys. Rev. E}, vol.~68, no.~2,
  p.~021502, 2003.

\bibitem{silbert2006structural}
L.~E. Silbert, A.~J. Liu, and S.~R. Nagel, ``Structural signatures of the
  unjamming transition at zero temperature,'' {\em Phys. Rev. E}, vol.~73,
  no.~4, p.~041304, 2006.

\bibitem{somfai2007critical}
E.~Somfai, M.~van Hecke, W.~G. Ellenbroek, K.~Shundyak, and W.~van Saarloos,
  ``Critical and noncritical jamming of frictional grains,'' {\em Phys. Rev.
  E}, vol.~75, no.~2, p.~020301, 2007.

\bibitem{van2009jamming}
M.~Van~Hecke, ``Jamming of soft particles: geometry, mechanics, scaling and
  isostaticity,'' {\em J. Phys. Condens. Matter}, p.~033101, 2009.

\bibitem{torquato2010jammed}
S.~Torquato and F.~H. Stillinger, ``Jammed hard-particle packings: From kepler
  to bernal and beyond,'' {\em Rev. Mod. Phys.}, p.~2633, 2010.

\bibitem{coulais2012dense}
C.~Coulais, {\em Dense vibrated granular media: from stuck liquids to soft
  solids}.
\newblock PhD thesis, Universit{\'e} Pierre et Marie Curie-Paris VI, 2012.

\bibitem{zaccone2011approximate}
A.~Zaccone and E.~Scossa-Romano, ``Approximate analytical description of the
  nonaffine response of amorphous solids,'' {\em Phys. Rev. E}, no.~18,
  p.~184205, 2011.

\bibitem{frenkel2001understanding}
D.~Frenkel and B.~Smit, {\em Understanding molecular simulation: from
  algorithms to applications}.
\newblock Academic press, 2001.

\bibitem{Note1}
See Supplemental Material for the extensive description of the generalized RDF.

\bibitem{milkus2016local}
R.~Milkus and A.~Zaccone, ``Local inversion-symmetry breaking controls the
  boson peak in glasses and crystals,'' {\em Phys. Rev. B}, vol.~93, no.~9,
  p.~094204, 2016.

\bibitem{Note2}
See Supplemental Material for a complete description of the 8 parameters fit.

\bibitem{wang2018microscopic}
D.~Wang, J.~Ren, J.~A. Dijksman, H.~Zheng, and R.~P. Behringer, ``Microscopic
  origins of shear jamming for 2d frictional grains,'' {\em Phys. Rev. Lett.},
  vol.~120, no.~20, p.~208004, 2018.

\bibitem{hartley2003logarithmic}
R.~Hartley and R.~Behringer, ``Logarithmic rate dependence of force networks in
  sheared granular materials,'' {\em Nature}, vol.~421, no.~6926, pp.~928--931,
  2003.

\bibitem{Note3}
See the Supplemental Material for a detailed explanation on origin of the weak
  forces.

\bibitem{hansen1990theory}
J.-P. Hansen and I.~McDonald, {\em Theory of simple liquids}.
\newblock Elsevier, 1990.

\end{thebibliography}


\begin{thebibliography}{0}%
\makeatletter
\providecommand \@ifxundefined [1]{%
 \@ifx{#1\undefined}
}%
\providecommand \@ifnum [1]{%
 \ifnum #1\expandafter \@firstoftwo
 \else \expandafter \@secondoftwo
 \fi
}%
\providecommand \@ifx [1]{%
 \ifx #1\expandafter \@firstoftwo
 \else \expandafter \@secondoftwo
 \fi
}%
\providecommand \natexlab [1]{#1}%
\providecommand \enquote  [1]{``#1''}%
\providecommand \bibnamefont  [1]{#1}%
\providecommand \bibfnamefont [1]{#1}%
\providecommand \citenamefont [1]{#1}%
\providecommand \href@noop [0]{\@secondoftwo}%
\providecommand \href [0]{\begingroup \@sanitize@url \@href}%
\providecommand \@href[1]{\@@startlink{#1}\@@href}%
\providecommand \@@href[1]{\endgroup#1\@@endlink}%
\providecommand \@sanitize@url [0]{\catcode `\\12\catcode `\$12\catcode
  `\&12\catcode `\#12\catcode `\^12\catcode `\_12\catcode `\%12\relax}%
\providecommand \@@startlink[1]{}%
\providecommand \@@endlink[0]{}%
\providecommand \url  [0]{\begingroup\@sanitize@url \@url }%
\providecommand \@url [1]{\endgroup\@href {#1}{\urlprefix }}%
\providecommand \urlprefix  [0]{URL }%
\providecommand \Eprint [0]{\href }%
\providecommand \doibase [0]{http://dx.doi.org/}%
\providecommand \selectlanguage [0]{\@gobble}%
\providecommand \bibinfo  [0]{\@secondoftwo}%
\providecommand \bibfield  [0]{\@secondoftwo}%
\providecommand \translation [1]{[#1]}%
\providecommand \BibitemOpen [0]{}%
\providecommand \bibitemStop [0]{}%
\providecommand \bibitemNoStop [0]{.\EOS\space}%
\providecommand \EOS [0]{\spacefactor3000\relax}%
\providecommand \BibitemShut  [1]{\csname bibitem#1\endcsname}%
\let\auto@bib@innerbib\@empty
\end{thebibliography}%

\end{document}



\title{Supplemental Material: Long-range anomalous decay of the correlation in jammed packings}


\author{Paolo Rissone}
\email{paolo.rissone@ub.edu}
\affiliation{Small Biosystems Lab, Department of Condensed Matter Physics, Carrer de Marti i Franques, 1, 11, 08028, Barcelona, Spain}

\author{Eric I. Corwin}
\email{ecorwin@uoregon.edu}
\affiliation{Department of Physics and Materials Science Institute, University of Oregon, Eugene, Oregon 97403, USA}

\author{Giorgio Parisi}
\email{giorgio.parisi@roma1.infn.it}
\affiliation{Dipartimento di Fisica, Sapienza Università di Roma, P.le Aldo Moro 5, 00185 Rome, Italy}
\affiliation{Istituto Nazionale di Fisica Nucleare, Sezione di Roma I, P.le A. Moro 5, 00185 Rome, Italy}
\affiliation{Institute of Nanotechnology (NANOTEC) - CNR, Rome unit, P.le A. Moro 5, 00185 Rome, Italy}

\date{\today}

\maketitle

\section*{Introduction to the general pair correlation function}

Let us first recall the equations introduced in the main text. Given a generic observable $\mathcal{O}$, let us define the correlation function between particles $i$ and $j$ at positions $\mathbf{r}_i$ and $\mathbf{r}_j$ respectively, by rewriting the RDF as the correlation function between particles $i$ and $j$ at positions $\mathbf{r}_i$ and $\mathbf{r}_j$ respectively, as
\begin{equation}
\label{eq:rdf_generic}
g^s_{\mathcal{O}}(r)= \frac{1}{C} \sum_{i,j} \delta(|\mathbf{r}_i - \mathbf{r}_j| - r) \mathcal{O}^s_i \mathcal{O}^s_j \, ,
\end{equation}
where $r$ is the distance between the particles $i$ and $j$, $C$ is the normalization factor and $s \in \Re$ a control parameter. Notice that for $s>0$ the main contribution to \eqref{eq:rdf_generic} comes from large values of $\mathcal{O}$, for $s<0$ the correlation between small values of $\mathcal{O}$ is enhanced and for $s=0$ the usual expression for the RDF, i.e. $g^{s=0}_{\mathcal{O}}(r) \equiv g(r)$, is recovered. However, \eqref{eq:rdf_generic} not only contains information about the point-to-point correlation of the observable $\mathcal{O}$ but also has a contribution due to the variation in the displacement of the particles through the system. This density-density correlation appears as periodic oscillations decreasing with the distance $r$. In order to suppress this added contribution let us define the \textit{generalized correlation function}
\begin{equation}
\label{eq:corr_generic}
C^s_{\mathcal{O}}(r) = \frac{g^s_{\mathcal{O}}(r)}{g(r)} \, ,
\end{equation}
where $g_{\mathcal{O}}^s(r)$ is given in \eqref{eq:rdf_generic} and $g(r)$ is the classic RDF accounting only for the density-density correlation. In this work we analyze the correlation of the forces between particles at jamming and the correlation of the variation of the coordination number w.r.t the isostaticity, $\Delta Z$.
%

%
\section*{RDF mid and long-range behavior}

We showed (see main text) that by introducing a system description w.r.t the contact points between particles it is possible to point out RDF features in the mid and long-range that are hidden by the usual system representation. To do this, let us note that the correlation decays with a behavior $\propto e^{-a r/\sigma '}$ with $a = 0.318 \pm 0.009$. 
The study of the function $g(r)e^{a r}$ evidenced that the RDF results by the superimposition of two different oscillating signals so that 
\begin{equation}
    \label{eq:RDFbehavior}
    g(r) \propto e^{-a r} \sum_{m,l} f_i(r) = e^{-a r} \sum_{m,l} c_i e^{-a_i r}\cos(p_i r + \psi_i) \,
\end{equation}
where $c_i, a_i, p_i, \psi_i$ are the function parameters and $m,l$ denote the oscillations in the mid and long-range, respectively. 
In Table \ref{tab:8paramsFIT} we reported the complete set of parameters obtained by fitting the RDF according to \eqref{eq:RDFbehavior}. Notice that 
$p_m \approx 2p_l = 7.514 \pm 0.003$ and $a_m \approx 2a_l =  0.70 \pm 0.01$ supporting the hypothesis that $f_l(r)$ can be considered as an $\mathcal{O}(2)$ correction to the $\mathcal{O}(1)$ mid-range leading term $f_m(r)$. 
Moreover, given that $c_m >> c_l$, the long-range oscillations get easily masked by the mid-range ones. Interestingly, the introduction of the fictive particles doubles the system size making it possible to observe both $f_m(r)$ and $f_l(r)$. This is not possible for the real particles system, where only the mid-range oscillations can be measured.

\setlength{\tabcolsep}{1.5mm}
\begin{table*}[h]
\centering
\begin{tabular}{ccccc}
  & $\mathbf{c_i}$ & $\mathbf{a+a_i}$ & $\mathbf{p_i}$ & $\mathbf{\psi_i}$\\ 
\toprule
$\mathbf{m}$ & 1.26 (4) & 0.70 (1) & 7.514 (3) & -7.68 (2)\\
$\mathbf{l}$ & -0.046 (1) & 0.31 (1) & 3.805 (3) & -8.00 (2)\\
\end{tabular}
\caption{\label{tab:8paramsFIT} Complete set of parameters obtained by fitting the contacts RDF to \eqref{eq:RDFbehavior}. $m,l$ denote the mid and long-range behavior, respectively. The error of each parameter is shown in parenthesis. Note that the total damping coefficients $a'_i = a + a_i$ as a consequence of the procedure adopted to analyze the $g(r)$.}
\end{table*}

%

%
\section*{Geometry of the forces network}
\begin{figure}[ht]
    \centering
    \includegraphics[width=0.3\textwidth]{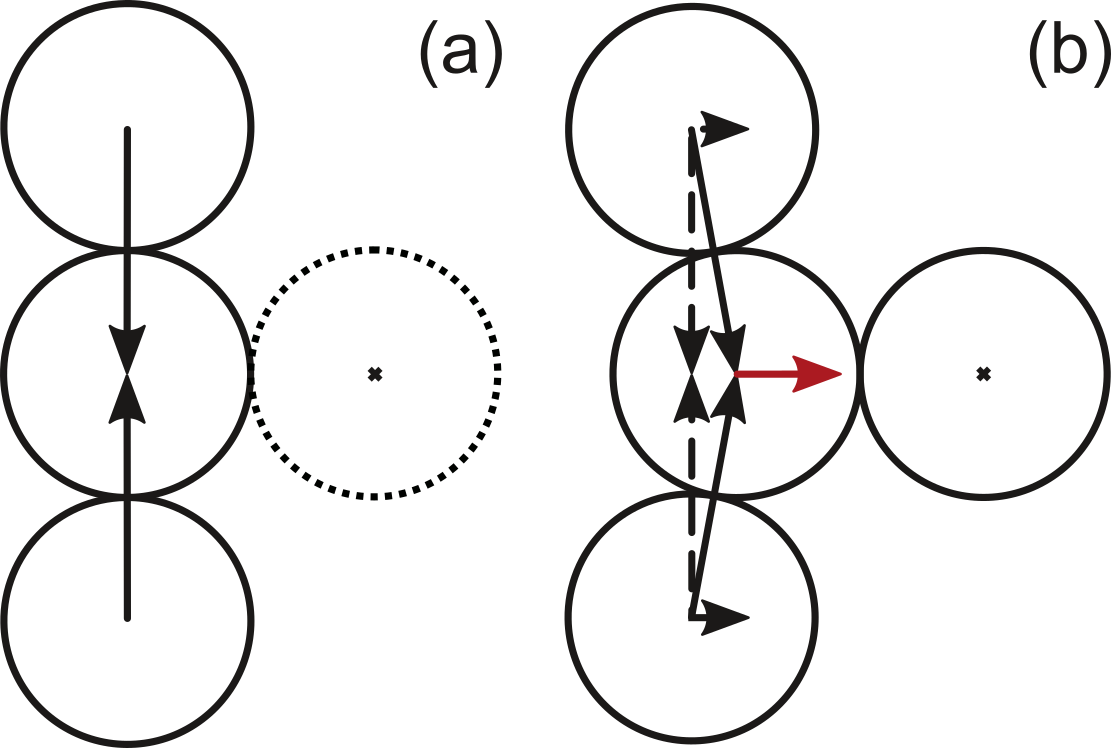}
    \caption{Schematic representation of the interparticle forces in $d=2$. \textbf{(a)} For particles lying in the same plane the forces are perfectly balanced (black arrows). No other particle is involved in the interaction and within this picture this is a strong force belonging to the main network. \textbf{(b)} When the particles are not coplanar, the extra components of the force constitute a weak interaction with another particle (red arrow).}
    \label{fig:forces_cartoon}
\end{figure}

When a system of monodisperse HS reaches the jamming point, the particles form a rigid network of contacts which is stabilized by the total balance of the forces exchanged in the contact points. Within this picture, is it possible to distinguish between \textit{strong} and \textit{weak} forces. In our work, we analyzed the weak forces originated by slight unbalances of the interactions between neighbouring particles. In Fig.\ref{fig:forces_cartoon} we schematically show the case of three HS in contact in $d=2$. When the particles lie on the same plane, i.e. are perfectly centrosymmetric, the exchanged force is coplanar and perfectly balanced so that no other particle can be part of the force network (Fig.\ref{fig:forces_cartoon}(a)). By contrast, if the middle sphere is out-of-plane the resulting force exhibits a small sideways component, which is the weak force (Fig.\ref{fig:forces_cartoon}(b)): the higher the coplanarity, the smaller the weak force.
